\documentstyle[pra,aps,twocolumn,psfig,epsf]{revtex}

\begin{document}

\draft

\title{Wave packet dynamics with Bose-Einstein condensates} 

\author{ R. Dum$^{1,2}$, A. Sanpera$^1$, K.-A. Suominen$^3$,  
\\
M. Brewczyk$^{4}$, M. Ku\'s$^5$, K. Rz\c a\.zewski$^5$, and M. Lewenstein$^1$}
\address{(1) Commissariat \`a l'Energie Atomique, 
DSM/DRECAM/SPAM, Centre d'Etudes de Saclay \\91191 Gif-sur-Yvette, France}
\address{
(2) Ecole Normale Sup\'erieure,
Laboratoire Kastler Brossel, 
24, Rue Lhomond, F-75231 Paris Cedex 05, France}
\address{(3)
Helsinki Institute of Physics, PL 9, FIN-00014 Helsingin yliopisto,
Finland}
\address{(4) Filia Uniwersytetu Warszawskiego,
ul. Lipowa 41, 15-424 Bia{\l}ystok, Poland}
\address{(5) Centrum Fizyki Teoretycznej PAN and College of Science,
 Al. Lotnik\'ow 32/46, 02-668 Warsaw, Poland}

\date{\today}

\maketitle

\begin{abstract}

We study wave packet dynamics of a Bose condensate in a
periodically shaken trap. Dichotomy, that is, 
dynamic splitting of the condensate, and dynamic stabilization are 
analyzed in analogy with similar phenomena 
in the domain of atoms in strong laser fields.

\end{abstract}


\narrowtext

\vspace{0mm}

Recently it has become possible to prepare  
Bose-Einstein condensates of alkali gases~\cite{BEC} with
up to $10^7$ magnetically
trapped atoms. The condensed state is a macroscopically populated
quantum state well localized in the magnetic trap.
It is, therefore, an ideal tool to study wave packet 
dynamics under experimentally feasible conditions.
There are many interesting quantum phenomena resulting from 
the electronic wave packet dynamics; in this Letter we refer, in particular, to the phenomena of {\it wave packet dichotomy and stabilization}
exhibited by an electron bound by an atomic 
potential in presence of a strong  laser field \cite{gavrila}. 
We argue that  the same phenomena occur
in the dynamics of the condensate wave function.
The analogy is based on the fact that 
in the frame of reference moving with the free  electron oscillating 
in the field, the Kramers-Henneberger frame, the effect of 
the laser  is equivalent to a periodic shaking of the atomic potential 
along the laser polarization axis. 
A condensate in a periodically 
shaken trap could, therefore, {\em a priori} show a similar behavior.

As the intense laser field drives  the 
electron, the process of ionization of the atom occurs.
By increasing the laser intensity, one normally increases 
the ionization rate. 
However, for very intense fields of high frequency, this rate  
eventually starts to decrease with  intensity -- 
this is called {\it atomic stabilization} \cite{gavrila}. 
In this process the electronic wave packet remains bounded, i.e.
well localized in space (without spreading),  although highly distorted due 
to the  combined effects of the laser field and the atomic potential .
This {\it effective} atom-laser potential exhibits a double well structure 
which splits the electronic wave packet into two spatially separate parts; 
this effect is called {\it dichotomy}.
To achieve stabilization it is necessary to turn on the laser adiabatically
in order to ensure that the atomic ground state will evolve 
to the ground state of this {\it effective} atom-laser potential.
This type of stabilization \cite{misha}
has never been observed experimentally \cite{FOM}, since
it requires very intense high frequency fields, 
which currently can only be generated  in a form of a very short pulse;  as
an electron in an atom is highly unstable,  it would thus be most
likely ionized during the  turn-on of such a pulse.

Let us analyze the analogy between the electron and the condensate in more detail. The electron bound by an atomic potential $U({\vec{r}})$ and interacting with a laser field of amplitude ${\cal E}{\vec{e}}_z$ (polarized along the $z$-direction) is, for our purposes, best described
%in the ``space-translated'' Schr\"odinger equation, i.e.
in the Kramers-Henneberger frame of reference, in which  the interaction
with the laser field results in an effective time dependent
``atomic'' potential:

\begin{eqnarray} \label{SILAP}
   \left[-i\hbar \partial_t
    -\frac{\hbar^2{\vec{\nabla}}^2}{2m_e}
  +U_e({\vec{r}}+\alpha_L\sin(\omega_L t)
{\vec{e}}_z)\right]\Psi_e({\vec{r}},t)=0,
\end{eqnarray}
where $\alpha_L= (e{\cal E})/(m_e\omega_L^2)$ is the electron excursion amplitude, while $\omega_L$ is the laser  frequency.

Consider now a condensate with $N$ atoms in a magnetic trap 
$V(\vec{r})$ which is periodically shaken along the $z$-axis.
In a Hartree-Fock treatment, the state of the condensate is described
by the Gross-Pitaevskii equation (GPE) \cite{BAYM}. It accurately describes the
wave function of the condensate $\Psi$ in presence of particle interactions in thermal equilibrium at temperatures well
below the critical temperature. 
Furthermore, the time dependent GPE describes the dynamics of
the condensate in more general time dependent 
conditions \cite{BAYM,EXCITATION,TH}. 
We, therefore, have:
\begin{equation}
  \left[-i\hbar\partial_{t}
       -\frac{\hbar^2 \vec{\nabla}^2}{2m}+
   V({\vec{r}}+\alpha(t){\vec {e}}_z) +g N|\Psi({\vec{r}},t)|^2\right] \Psi({\vec{r}},t)=0. \label{GPE}
\end{equation}
Here $\alpha(t)=\alpha_0\sin(\omega t)$
is the shaking  amplitude. 
If we identify the amplitude $\alpha_0$ and the frequency $\omega$ of the
shaking with
the electron excursion amplitude 
$\alpha_L$ and the laser frequency 
$\omega_L$ respectively,
 Eq.(\ref{GPE}) is very similar to Eq.(\ref{SILAP}).
An important difference is the presence of a nonlinear coupling term with a coupling constant $g=4\pi\hbar^2a_s/m$, where  $a_s>0$ is  the
$s$-wave scattering length. 
The effects due to the presence of the nonlinear term as well 
as the larger mass of the atom will be discussed below.
We model the trapping potential by
\begin{equation}\label{MODEL}
   V({\vec{r}})= m(\Omega_x^2 x^2+\Omega_y^2 y^2+\Omega_z^2 z^2)/2,
\end{equation}
for $V({\vec{r}}) < V_c$,  and $V({\vec{r}})=V_c$ otherwise. This is a harmonic
potential with frequencies $\Omega_{x,y,z}$ which is cut at an energy $V_c$.
It thus resembles $U_e({\vec{r}})$,
in particular we have that
$V({\vec{r}}\to\infty)$ like $U_e({\vec{r}}\to \infty)$ is constant.
This trapping potential can be realized as  an
effective (adiabatic) trap potential ``dressed'' by the microwave coupling
between a trapped and an untrapped state. We will discuss this model later on in the paper.
Another possible realization would be a
condensate in a dipole trap potential formed by a strong off-resonant laser field.
The condensation might take place in such a trap, or the magnetically trapped
condensate may be loaded into it. 
Note that in absence of the cut $V_c$, i.e. for an exactly harmonic potential
the shaking of the trap 
would lead to an undistorted oscillation of the condensate wave function.

If  the time scale 
$1/\omega_L$ in Eq.(\ref{SILAP}) is shorter than the typical time scale
of the electronic motion,  the combined laser-atom potential can be replaced by its  time average over one laser period; 
ideally the wave function of the electron will then evolve adiabatically from a state in the 
bare atomic potential into
the corresponding state of this time-averaged atom-laser potential \cite{gavrila}.
For a large enough amplitude $\alpha_L$, this time average potential will have 
a double well structure and thus the electron wave function will exhibit dichotomy.
Using the same reasoning for the condensate 
we replace the GPE (Eq.(\ref{GPE})) by a
GPE with a time-averaged potential
if the time scale $\sim 1/\omega$ of the shaking 
is shorter than all other relevant time scales: 
\begin{equation}
 \left[-i \hbar\partial_t
   -\frac{\hbar^2 \vec{\nabla}^2}{2m}+
      V_{\rm eff}(\vec{r},\alpha_0)
   + gN|\Psi(\vec{r},t)|^2 \right]\Psi(\vec{r},t)=0 \label{GPEAV},
\end{equation}
where the time-averaged potential is given by
\begin{equation}
   V_{\rm eff}(\vec{r},\alpha_0)=
\frac{1}{2\pi}\int_{-\pi}^{\pi} d\varphi V(\vec{r}+\alpha_0\sin(\varphi) {\bf e}_z);
   \,\, \varphi=\omega t.
\end{equation}

For sufficiently large shaking amplitudes, a double well structure
appears as illustrated  in Fig.~1(a).
We therefore expect dichotomy as well as stabilization to appear also,
due to the trapping of the atomic wave function in the two potential wells of $V_{\rm eff}$ near the turning points of the oscillation. 
Indeed, for large enough $\alpha_0$, stable states are dynamically 
formed in the double well. 
Even though the particles in such states are 
sometimes out of the trap, 
no ``ionization" occurs. This is due to the fact that the time scale for
particle motion is much larger than the time scale $1/\omega$ of shaking, 
so that the particles do not have enough time to react on being momentarily 
out of the trap. Due to the large atomic mass, the 
requirements on $\omega$ are in fact much less stringent than in the electronic case.

The initial state of the condensate for temperature
$T\simeq 0$
is given by the lowest eigenstate $\Psi_0$ of 
the time independent GPE
\begin{equation}\label{GPE0}
[-\mu-{\hbar ^2 \vec{\nabla}^2\over 2m} +
V({\vec{r}})+Ng|\Psi_0({\vec{r}})|^2]\Psi_0({\vec{r}})=0.
\end{equation}
where $\mu$ is the chemical potential.
By replacing $V$ by $V_{\rm eff}$ in Eq.(\ref{GPE0}) we determine the ground state of
the system in presence of the double well potential. The corresponding
chemical potential is $\mu_{\rm eff}$.

\begin{figure}[h]
\centerline{
\vspace*{0.4cm}
\psfig{height=90mm,width=72mm,file=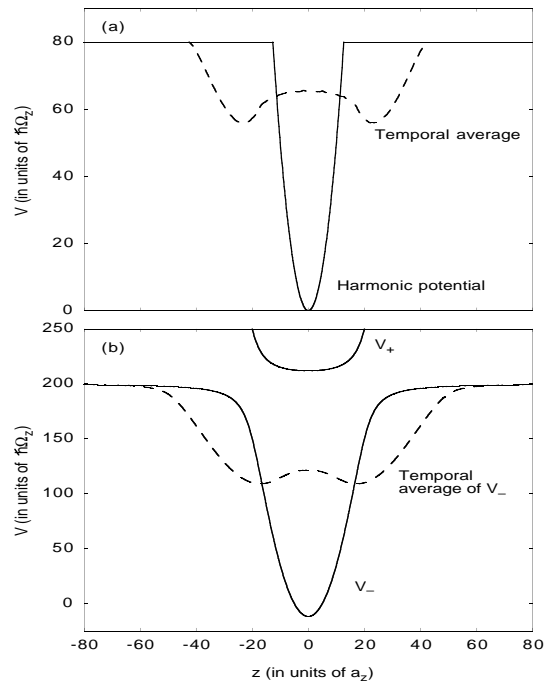}
}
\caption[f1]{(a) Cut harmonic potential(solid line) and time averaged potential 
(dashed line) for $\alpha_0=30 a_z$ and $V_c=80\hbar\Omega_z$; 
$a_z=(\hbar/2 m\Omega_z)^{1/2}$ and $gN=100$.
(b) Dressed state potentials  ($V_+$ and $V_-$) 
and time averaged potential of the lower one ($V_-$) from Eq.(\ref{dressed}); 
$\alpha_0=30 a_z$, 
$\omega_R=100\Omega_z$, $\Delta=200\Omega_z$.}
\label{figures1}
\end{figure}

In contrast to Eq.(\ref{SILAP}), the GPE accounts for atomic interactions, 
so that the condensate splitting depends on $g N$. With increasing $N$ the wave function will at some point overcome the potential barrier between the wells in $V_{\rm eff}$, and the splitting will disappear.
This will happen when the effective chemical potential ($\mu_{\rm eff}$) in the double well potential $V_{\rm eff}$ exceeds the height of the
double well. Therefore the number $N$ is limited from above.

The validity of the above picture depends on the turn-on time
of the shaking. Ideally, an adiabatic switching will transfer
the ground state of the GPE (Eq.(\ref{GPE0})) at $t=0$ to the ground state of the
GPE with $V$ replaced $V_{\rm eff}$.
Too rapid passage  from the 
cut-harmonic potential into the double well form induces transitions to the 
higher states of $V_{\rm eff}$ and even into the continuum. 
To avoid these transitions, we turn on the shaking gradually,
setting
\begin{equation} \label{adiab}
   \alpha(t)=
   \cases{\alpha_0\sin^2      \left(\frac{\pi}{2}\frac{t}{t_{\rm on}}\right)\sin(\omega t), & for $0\le t\le t_{\rm on}$\cr
\cr
\alpha_0 \sin(\omega t) , & for $t\ge t_{\rm on}$\cr}
\end{equation}

The turn-on time $t_{\rm on}$ must be  $\gg 2\pi/\omega$; in practice we take
$t_{\rm on} \ge  (50-150)\times 2\pi /\omega\simeq 2\pi\alpha_0/a_z\omega$, where $a_z=\sqrt{\hbar/2 m\Omega_z}$.

A typical experimental sequence to demonstrate the above phenomena  
will consist in the following.
Initially,  the system is prepared  in the state $\Psi_0$ of Eq.(\ref{GPE0}).
At time $t=0$ we slowly start shaking the trap along
the $z$-axis (Eq.(\ref{adiab})).
Let us first  assume that the perpendicular motion does not 
play a significant role, 
and consider the motion along the $z$-axis only.
In  Fig. 2(a) we show the the time evolution of the 1D-condensate wave function. The parameters used are the same as for Fig. 1(a), 
i.e. $\omega=10\Omega_z$, $V_c=80 \hbar\Omega_z$, $\alpha_0=30a_z$, and $gN=100$. 
The gradual splitting of the condensate wave function can be clearly seen.
Moreover almost all particles remain trapped.

Due to the cut $V_c$ in the trapping potential (Eq.(\ref{MODEL})) atoms 
may escape from the trap. However, as can bee seen from Fig. 2(a), almost no atoms do it; this indicates stabilization.
To study this phenomenon in more detail we  lower
the cut-off energy $V_c$ to $50 \hbar \Omega_z$ to favor the escape 
from the trap. The escape rate is calculated using standard absorbing boundary conditions \cite{gavrila}(b); we observe a {\it decrease} in the escape rate 
if the shaking amplitude $\alpha_0$, {\it increases}. More specifically, increasing $\alpha_0$ from 15$a_z$ to  20$a_z$,
the escape rate decreases by a factor $\simeq 2$. 
Due to the large atomic mass the escape rate is overall very small, i.e.
$\leq 1\%$ of the trapped population per 100 shaking cycles.
For the same reason, the condensate stabilization occurs already 
for relatively small shaking frequencies $\hbar\omega<V_c-\mu_{\rm eff}$. 
The condensate stabilization is, therefore, more pronounced and
can be achieved in experimentally more accessible conditions than its analog in strong field  dynamics of electrons.

\begin{figure}[h]
\centerline{
\psfig{height=58mm,width=72mm,file=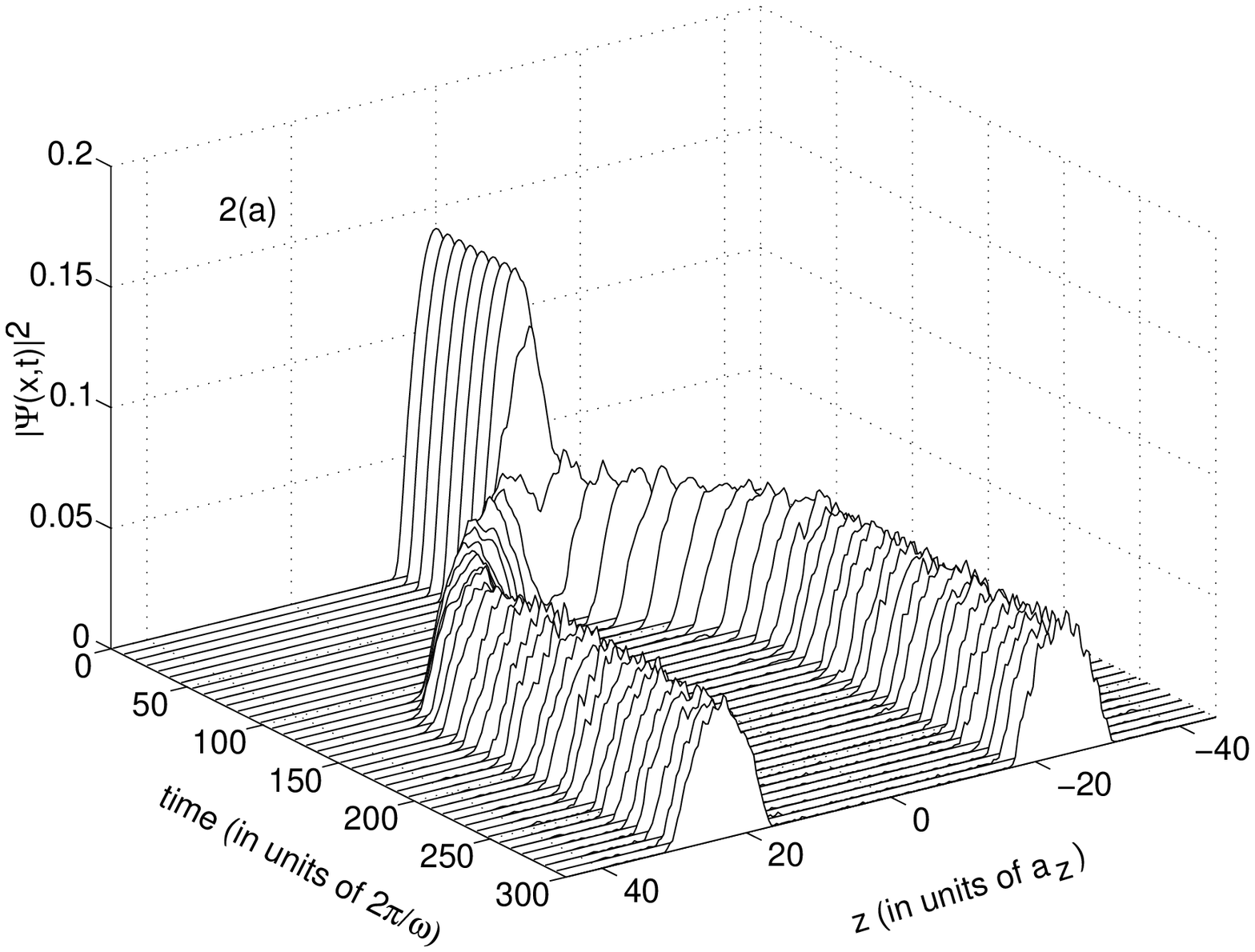}
}
\end{figure}

\begin{figure}[h]
\centerline{
\vspace*{0.3cm}
\psfig{height=58mm,width=72mm,file=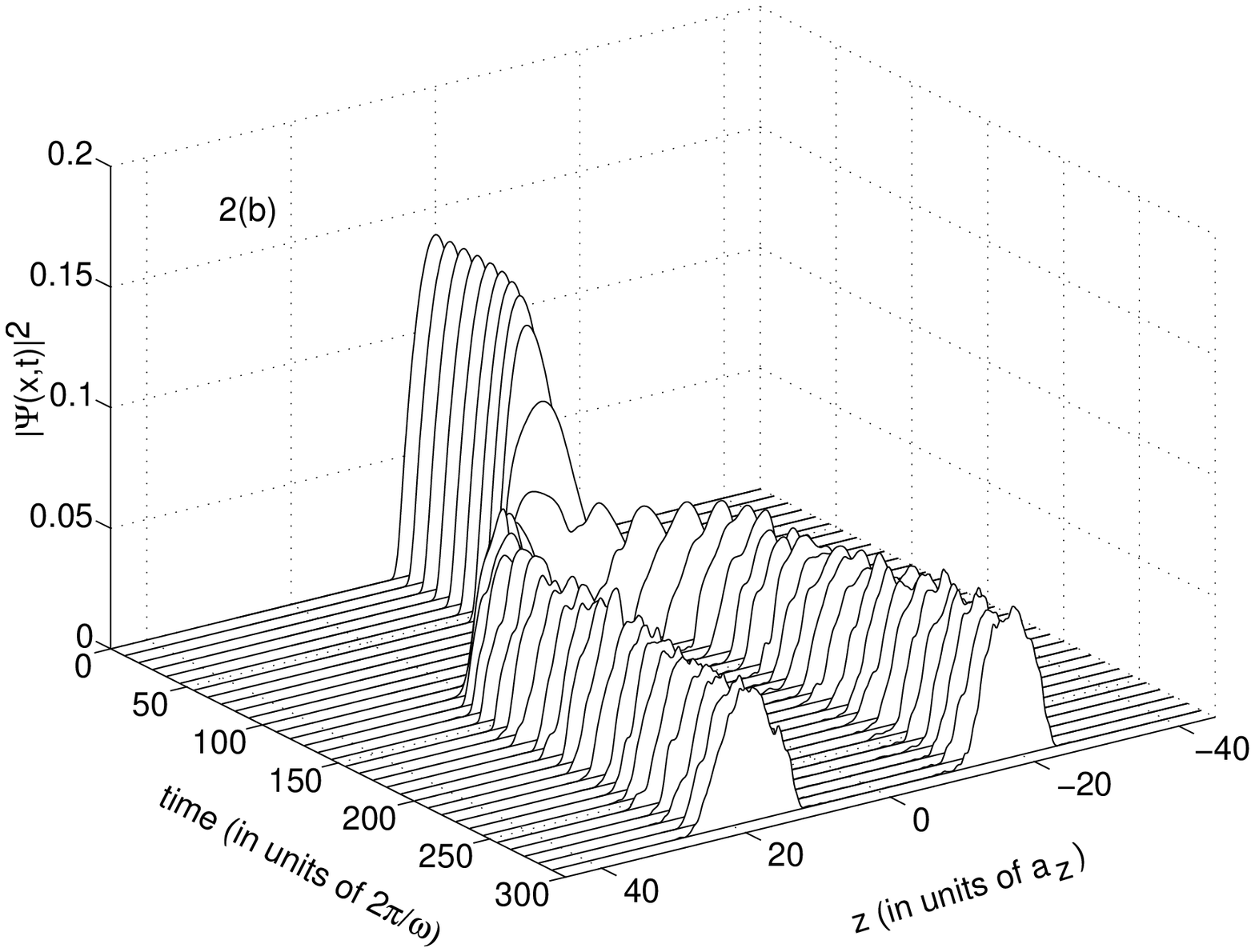}
}
\caption[f2]{(a) Time evolution of the condensate density $|\Psi(z,t)|^2$
undergoing 300 shaking cycles; $\alpha_0=30a_z$, $\omega=10\Omega_z$, $t_{\rm on}=150\times(2\pi/\omega)$,
$V_c=80\hbar\Omega_z$. The nonlinear coupling is $gN=100$, that
corresponds to $\hbar\mu = 14.13\Omega_z$; (b) Same as (a) for
the total condensate density $|\Psi(z,t)|^2=
|\Psi_1(z,t)|^2+|\Psi_0(z,t)|^2$, 
calculated from the 1D two state model; 
$\alpha_0=30 a_z$, $\omega=2.5\Omega_z$, $t_{\rm on}=150\times(2\pi/\omega)$,
$\omega_R=100\Omega_z$,\,$\Delta=200 \Omega_z$.}
\label{figure2}
\end{figure}

We turn now to a discussion of a model with two internal states, a trapped  state
$(F,m_F\ne 0)$ and an untrapped state $(F',m_F=0)$ .
They are coupled via a microwave field which allows coherent transitions between the states
\cite{MITalaser}.
In the rotating wave approximation the GPE is given by 
\begin{eqnarray} \label{TWO}
&& \left[
   -i\hbar\partial_{t} -\frac{\hbar^2 \nabla^2}{2m}
    +g N (|\Psi_0({\bf r},t)|^2+|\Psi_1({\bf r},t)|^2) \right. \nonumber \\
&&+ 
\left.
\left(
\begin{array}{cc}
m\Omega_z^2 (z+\alpha(t))^2/2 & \hbar \omega_R/2 \\
\hbar \omega_R/2 & \hbar \Delta
\end{array}
\right)
\right]
\left(
\begin{array}{l}
\Psi_0({\bf r},t \\
\Psi_1({\bf r},t)
\end{array}
\right)=0.
\end{eqnarray}
Here $\Psi_0, \Psi_1$ are the wave functions of atoms in the trapped and untrapped state
 normalized to the respective fraction of atoms in these states,
$\Delta$ is the detuning of the microwave from the transition frequency,
and $\omega_R$ is the Rabi frequency of the microwave transition. 
For simplicity we assume all coupling constants equal to $g$. 
For a large enough Rabi frequency $\omega_R$, the coupled states can be replaced 
by uncoupled dressed states with energies:

\begin{eqnarray}
V_{\pm}(&z&,t)=\frac{1}{2} \left[
m\Omega_z^2 ( z+\alpha(t))^2/2 +\hbar\Delta \right. \nonumber\\
&\pm& \left.\sqrt{ \left( m\Omega_z^2 (z+\alpha(t))^2/2 -\hbar\Delta\right)  ^2+\hbar^2\omega_{R}^2 }\,\,\,\,\right]\, .\label{dressed}
\end{eqnarray}

The potential $V_-$ resembles our model potential;
this is illustrated in Fig. 1(b) in which we also plot  
time averages of the  time dependent dressed 
state potential.
Obviously Eq.(\ref{GPE}), with $V$ replaced by $V_-$,  is not exact, 
since it neglects entirely the non-adiabatic transitions to an upper branch of the 
dressed potential. The model is nevertheless reasonable, since the
exact numerical treatment based of the 
two--component GPE  leads to very similar results,  as shown in  Fig.~2(b).
In the model of Eq.(\ref{TWO}) we account rigorously for non-adiabatic (Landau-Zener) 
transitions from the lower to the upper ``dressed'' state manifold. 
Obviously,
some of the atoms may cease to oscillate in the lower potential and
instead they may be transferred to the upper  ``dressed'' potential, destroying the dichotomy.
In order to avoid the Landau-Zener transitions $\omega$
has to be smaller than a certain critical value \cite{adiab}.
On the other hand for the validity of the time-averaged potential
$\omega$ has a lower limit of approximately $2 \Omega_z$.
In consequence dichotomy and stabilization in a realistic system occur in a
limited range of $\omega$'s.

Finally,  we have  generalized our study
to a 3D case.
We chose parameters
that resemble those of the  MIT experiment \cite{BEC}, that is a cigar shaped trap
with a small $\Omega_z$ and equal frequencies perpendicular to it. We assume the shaking
occurs along the long z-axis.
As shown in  Fig.~3 the presence of perpendicular motion does not invalidate
the conclusions from the 1D approximation: 
the splitting of the condensate is clearly visible. 

\begin{figure}[h]
\centerline{
\vspace*{0.5cm}
\psfig{height=58mm,width=72mm,file=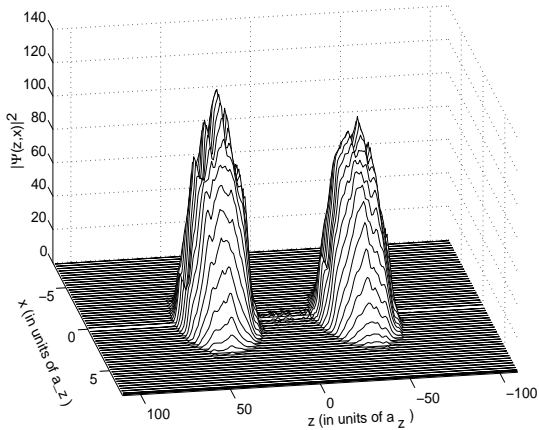}
}
\caption[f3]{2D cross--section 
of the condensate density $|\Psi(x,z, t)|^2$ for a cigar shaped trap at
$t=400(2\pi/\omega)$; $\alpha_0=60a_z$, $\omega=10\Omega_z$, $t_{\rm on}=250\times (2\pi/\omega)$,
$V_c=30\hbar\Omega_z$; $gN=100$, and $\Omega_x=\Omega_y=5\Omega_z$.
\label{figure3}}
\end{figure}

One should mention that the completeness of our analysis requires 
to check the stability of
the solutions of the time dependent GPE. Unstable solutions might lead to a depletion of the condensate \cite{STABLE}. In a first
attempt we verified that the solutions of an oscillating condensate in a harmonic trap
lead to stable solutions of the GPE,  and therefore to no depletion. 

Summarizing, we have shown that  
Bose-Einstein condensates are
ideal tools to  study wave packet behavior first
predicted  in the realm of atoms in superstrong laser fields.
We have shown that both, dichotomy and stabilization against atom escape
can be achieved in condenstates.
Although these effects  are very difficult to
realize  in electron-atom systems, their
observation in condensates seems quite feasible.
Furthermore, we believe that wave packet dynamics of condensates might lead to
interesting possibilities  of condensate state engineering.
For instance, so far double peaked condensates  have been created using laser ``knives'' 
that cut a single  condensate into two parts.
We offer here an alternative method to achieve a similar dichotomy in a
more controlled way which  opens new perspectives for condensate interference
studies.

K.-A. S. and A. S. acknowledge the financial support from the Academy of Finland.
This work was supported by Project de Cooperation France-Pologne 6423.
M.B. and K.R. acknowledge the support of KBN Grant No. 2P03B04209.


\begin{references}
\bibitem{adiab} The Landau-Zener probability for a nonadiabatic transition is
$\exp(-2\pi\Lambda); \Lambda=\hbar(\omega_R/2)^2/(v\lambda)$, where
$v=\alpha_0\omega$ is the atomic speed at the resonance point $z_0$
($m\Omega_z^2z_0^2/2=\hbar\Delta$) and $\lambda=|\partial
(m\Omega_z^2z^2/2-\hbar\Delta)/\partial z|_{z=z_0}$. The condition for
adiabatic following is $\Lambda\gg 1$, which leads to $\alpha_0\omega\ll
\omega_R^2(m\Omega_z^2\Delta/\hbar)^{-1/2}$.
\end{references}
\end{document}